\documentclass{PoS}

\usepackage[percent]{overpic}
\usepackage{color}
\usepackage{amsmath}
\usepackage{amssymb}
\usepackage{mathrsfs}
\usepackage{siunitx}
\usepackage{booktabs}
\usepackage{fancybox}
\DeclareMathAlphabet{\mathcal}{OMS}{cmsy}{m}{n}
\usepackage[utf8]{inputenc} 
\usepackage{booktabs}

\title{Searches for point-like sources of cosmic neutrinos with 11 years of ANTARES data}

\ShortTitle{Searches for point-sources of cosmic neutrinos with 11 years of ANTARES data}

\author{
The ANTARES Collaboration\footnote{For collaboration list, see PoS(ICRC2019) 1177.}\\
\itshape \href{http://antares.in2p3.fr/Collaboration/index2.html}{http://antares.in2p3.fr/Collaboration/index2.html}\\
E-mail: \email{julien.aublin@apc.in2p3.fr, giulia.illuminati@ific.uv.es, sergio.navas@ugr.es}
}

\abstract{

The main goal of the ANTARES neutrino telescope is the identification of neutrinos from cosmic accelerators. The good visibility towards the Southern sky for neutrino energies below 100 TeV and the good angular resolution for reconstructed events make the telescope excellent to test for the presence of point-like sources, especially of Galactic origin. The median angular resolution for track-like events (mainly from $\nu_{\mu}$ CC interactions) is $0.4^{\circ}$ while the median angular resolution for contained shower-like events (mainly from $\nu_{e}$ CC and all-flavour NC interactions) is $3^{\circ}$. Recently the ANTARES Collaboration published the result of the search for cosmic point-like neutrino sources using track-like and shower-like events collected during nine years of data taking. In this contribution, an update to this analysis using eleven years of data recorded between early 2007 and the end of 2017, for a total livetime of 3136 days, is presented. Moreover, the results of a search for time and space correlation between the ANTARES events and 54 IceCube tracks and those of the searches for neutrino candidates associated with the IceCube-170922A event or from the direction of the TXS 0506+056 blazar are reported.

\vspace{4mm}
{\bfseries Corresponding authors:}
Julien Aublin$^{1}$, Giulia Illuminati$^{2}$, \speaker{Sergio Navas}$^{3}$\\
{$^{1}$ \itshape APC, Univ Paris Diderot, CNRS/IN2P3, CEA/Irfu, Obs de Paris, Sorbonne Paris Cit\'e, France}\\
{$^{2}$ \itshape IFIC - Instituto de F\'isica Corpuscular (CSIC - Universitat de Val\`encia) c/ Catedr\'atico Jos\'e Beltr\'an, 2 E-46980 Paterna, Valencia, Spain}\\
{$^{3}$ \itshape Dpto. de F\'\i{}sica Te\'orica y del Cosmos \& C.A.F.P.E., University of Granada, 18071 Granada, Spain}\\
}

\FullConference{36th International Cosmic Ray Conference -ICRC2019-\\
		July 24th - August 1st, 2019\\
		Madison, WI, U.S.A.}

\begin{document}

\section{Introduction}\label{sec:intro}
The search for astrophysical neutrinos in the TeV-PeV range is among the primary goals of the ANTARES neutrino telescope~\cite{ANTARESdetector}. Cosmic neutrinos are expected to be produced through the decay of charged mesons, results of hadronic interactions of cosmic rays with matter or radiation in the surroundings of the acceleration sites. Unlike charged cosmic rays, neutrinos are not deflected by cosmic magnetic fields, thus allowing precise pointing to their production sites. Moreover, in contrast to gamma rays, neutrinos, being weakly interacting particles, can escape dense celestial environments, providing insight into the interior of the sources. The recent highly significant observations of an isotropic high-energy cosmic flux reported by the IceCube Collaboration \cite{IC3years, IC4yproc, IC6yproc, IC-VHE-29, IC-VHE-36}, followed by the first evidence of neutrino emission from an individual source, the blazar TXS 0506+056 \cite{ICTXS}, represented a major breakthrough in the field of neutrino astronomy and strongly motivate further investigations. The ANTARES neutrino telescope, located in the Northern hemisphere, with a clear visibility of the Galactic Plane, and with a very good angular resolution, provides an excellent tool for the search for point-sources, especially of Galactic origin. Here, the results of the latest searches for point-like sources using the ANTARES neutrino telescope are presented. The analysis  includes both track-like and shower-like events recorded in ANTARES between January 29, 2007 and December 31, 2017, for a total livetime of 3125.4 days. Track-like events are mainly originated by the passage in water of relativistic muons produced in charged current (CC) interactions of muon neutrinos. Shower-like events are mainly induced by neutral current (NC) interactions, and by ${\nu_e}$ and ${\nu_\tau}$ CC interactions. The final sample employed in this analysis includes 8754 tracks and 195 showers. The events are selected following the chain of cuts applied on parameters provided by the reconstruction algorithms defined in \cite{lastPSant}. The selection criteria were optimised to minimise the neutrino flux needed for a $5\sigma$ discovery of a point-like source emitting with a $\Phi(E) \propto E^{-2.0}$ spectrum. 
The selected tracks are reconstructed with a median angular resolution better than $0.4^{\circ}$ for energies above 100 TeV, while a median angular accuracy of $\sim$3$^{\circ}$ is achieved for showers.

\section{Search Method}\label{sec:method}

An unbinned method based on an extended maximum likelihood ratio  test statistic is employed to identify clusters of cosmic neutrinos from point-like sources over the background of randomly distributed atmospheric background. The used likelihood is defined as

\begin{align} \label{eq:pslik}
    \log \mathscr L = \sum_{j} \sum_{i\in j} \log \Big[ \mu_\mathrm{sig}^{j}
     \mathcal{S}^{j}_{i} + {\mathcal N}^{\,j}  {\mathcal B}^{\,j}_{i} \Big]
    - \mu_\mathrm{sig} .
\end{align}

In this equation, $j$ denotes the sample ($\it{tr}$ for tracks, $\it{sh}$ for showers), $i$ indicates the event of the sample $j$, $\mu_\mathrm{sig}^{j}$ is the number of signal events fitted in the $j$ sample, ${\mathcal N}^{j}$ is the total number of events in the $j$ sample, $\mathcal{S}^{j}_{i}$ and ${\mathcal B}^{\,j}_{i}$ are the values of the signal and background PDFs for the event $i$ in the sample $j$, and $\mu_\mathrm{sig} = {\mu}^{tr}_\mathrm{sig} + {\mu}^{sh}_\mathrm{sig}$ is the total number of fitted signal events.

The signal and background PDFs are given by the product of a directional and an energy term. A parameterization of the point spread function, i.e. the probability density function of reconstructing an event at a given angular distance from the true source location, is employed as spatial PDF for the signal. For the background, the observed declination distribution of the selected data events is used. The energy PDFs are given by the probability density function of the energy estimator for signal, derived from Monte Carlo simulations of $E^{-\gamma}$ energy spectrum cosmic neutrinos, and background, obtained from simulations of atmospheric neutrinos using the spectrum of \cite{atmonu}.
When a search for point sources with time-dependent fluxes is performed, a time-dependent term is incorporated into the signal and background PDFs. The shape of the signal time-dependent term is either provided by observations of other experiments or assumed as generic Gaussian profile (see Section \ref{sec:search}). Given the small expected contribution of a cosmic signal in the overall data set, the time PDF for background is built using the time distribution of data events, ensuring a time profile proportional to the measured data.

The number of signal events $\mu^\mathrm{tr}_\mathrm{sig}$ and $\mu^\mathrm{sh}_\mathrm{sig}$ are fitted in the likelihood maximization. Moreover, the position in the sky of the source is either kept fixed or allowed to be fitted within specific limits depending on the type of search (see Section \ref{sec:search}). In case of time-dependent searches with generic Gaussian time PDF, the duration of the transient source emission is an additional free parameter of the likelihood.

The signal likeness of a cluster is determined by a test statistic computed as

\begin{equation}
    \mathcal Q = \log \mathscr L^{\rm {max}} - \log \mathscr L^{\rm {bkg}},
    \label{eq:teststat}
\end{equation}

where $\mathscr L^{\rm {max}}$ and $\mathscr L^{\rm {bkg}}$ are the values of the likelihood defined in equation~\ref{eq:pslik} calculated using the best-fit values of the free parameters and for the background-only case (${\mu}^\mathrm{tr}_\mathrm{sig} = {\mu}^\mathrm{sh}_\mathrm{sig} = 0$), respectively.

In order to calculate the significance of any observation, the observed $Q$ is compared to the test statistic distribution obtained in background-only pseudo-experiments (PEs) -- pseudo-data sets of data randomised in time. The fraction of $Q$ values which are larger than the observed $Q$ gives the significance (p-value) of the observation.

\section{Searches and Results}\label{sec:search}

Three different searches for steady astrophysical neutrino sources are performed: a scan over the whole ANTARES visible sky, a survey of 112 astrophysical candidates and an investigation of 75 IceCube tracks. A subset of the selected 75 IceCube candidates was recently investigated by ANTARES in a time-dependent analysis to search for a possible transient origin of the IceCube tracks \cite{timecorr}. These approaches are described below together with the corresponding results. Moreover, the results of dedicated searches for cosmic neutrinos associated either with the IC170922A event or with the TXS 0506+056 blazar are reported.

\subsection{Full-Sky Search}
\label{fullsky}
\begin{figure*}[ht]
\centering
\begin{overpic}[width=0.95\textwidth]{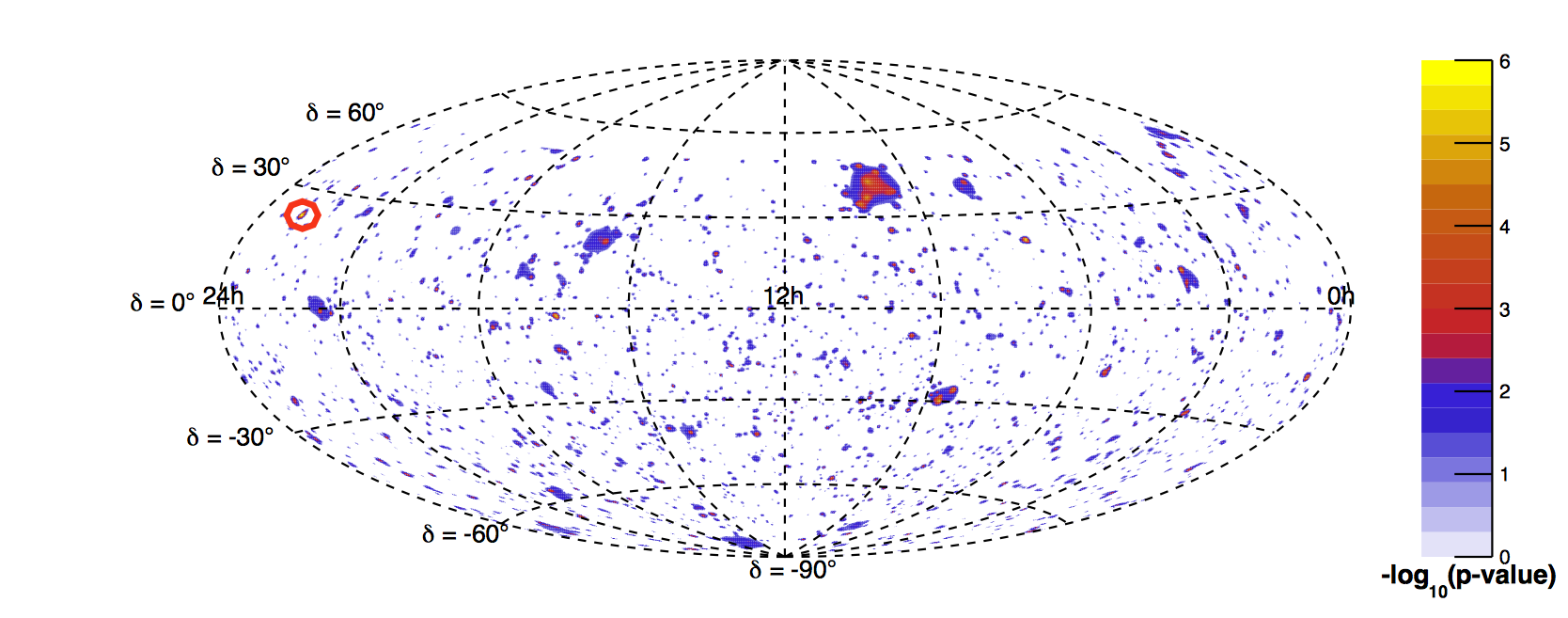}
     \put (66,2.7){\color{black} PRELIMINARY}
    \end{overpic}

\caption{Sky map in equatorial coordinates of pre-trial p-values for a point-like source of the ANTARES visible sky. The red contour indicates the location of the most significant cluster of the full-sky search.}
\label{fig:MapFSS}
\end{figure*}

In the full-sky search, an excess of signal events located anywhere in the ANTARES visible sky is searched for, without making any assumption about the source position. To this purpose, the $\mathcal Q$-value defined in equation~(\ref{eq:teststat}) is evaluated in steps of $1^\circ \times 1^\circ$ over the whole scanned region, with the location of the fitted cluster being left free to vary within these boundaries. The most significant cluster of this search, i.e. the cluster with lowest pre-trial p-value, is found at a right ascension of $\alpha = 343.7^\circ$ and a declination of $\delta = 23.6^\circ$ with a pre-trial p-value of $\num{1.5e-6}$. Figure~\ref{fig:MapFSS} shows the position of the cluster and the pre-trial p-values for all the directions in the ANTARES visible sky.
The post-trial significance of the cluster, obtained by comparing the pre-trial p-value to the distribution of the smallest p-values found anywhere in the sky when performing the same analysis on many PEs, is $23\,\%$ ($1.2\sigma$). The distribution of ANTARES events around the best-fit location of the cluster is shown in Figure~\ref{fig:Clusters}-left. It contains 18 (3) tracks within $5^\circ (1^\circ)$ and 1 shower event within $5^\circ$.

\subsection{Candidate List Searches}\label{candidate}
In the candidate list search, the directions of a pre-selected list of potential neutrino sources are investigated to look for an excess of neutrino events or, in case of null observation, to determine an upper limit on their neutrino fluxes. The description of the candidate list searches performed in this analysis follows.

\begin{figure*}[ht]
\centering
\begin{overpic}[width=0.49\textwidth]{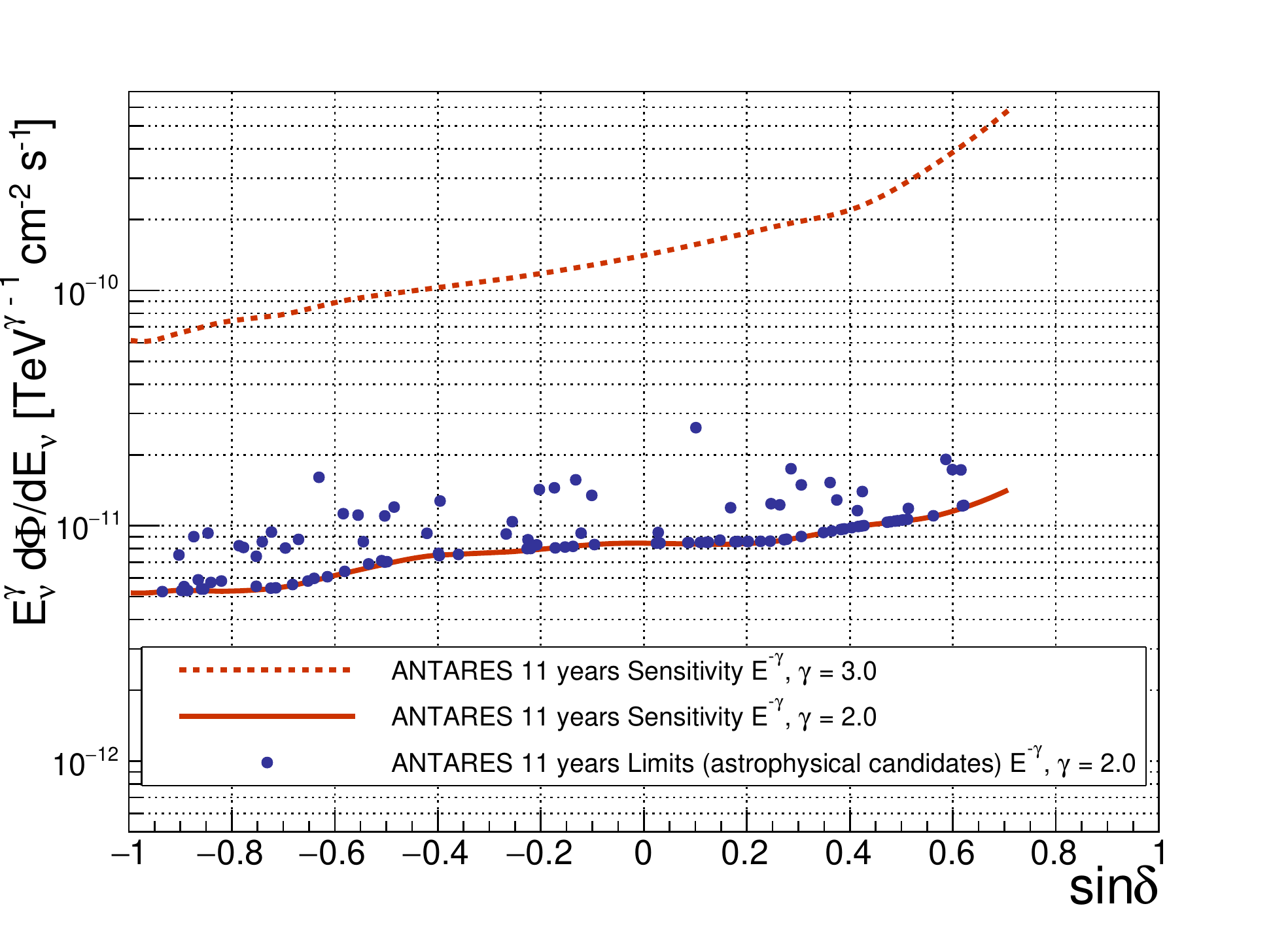}
     \put (11,60){\color{black} PRELIMINARY}
    \end{overpic}
    \begin{overpic}[width=0.49\textwidth]{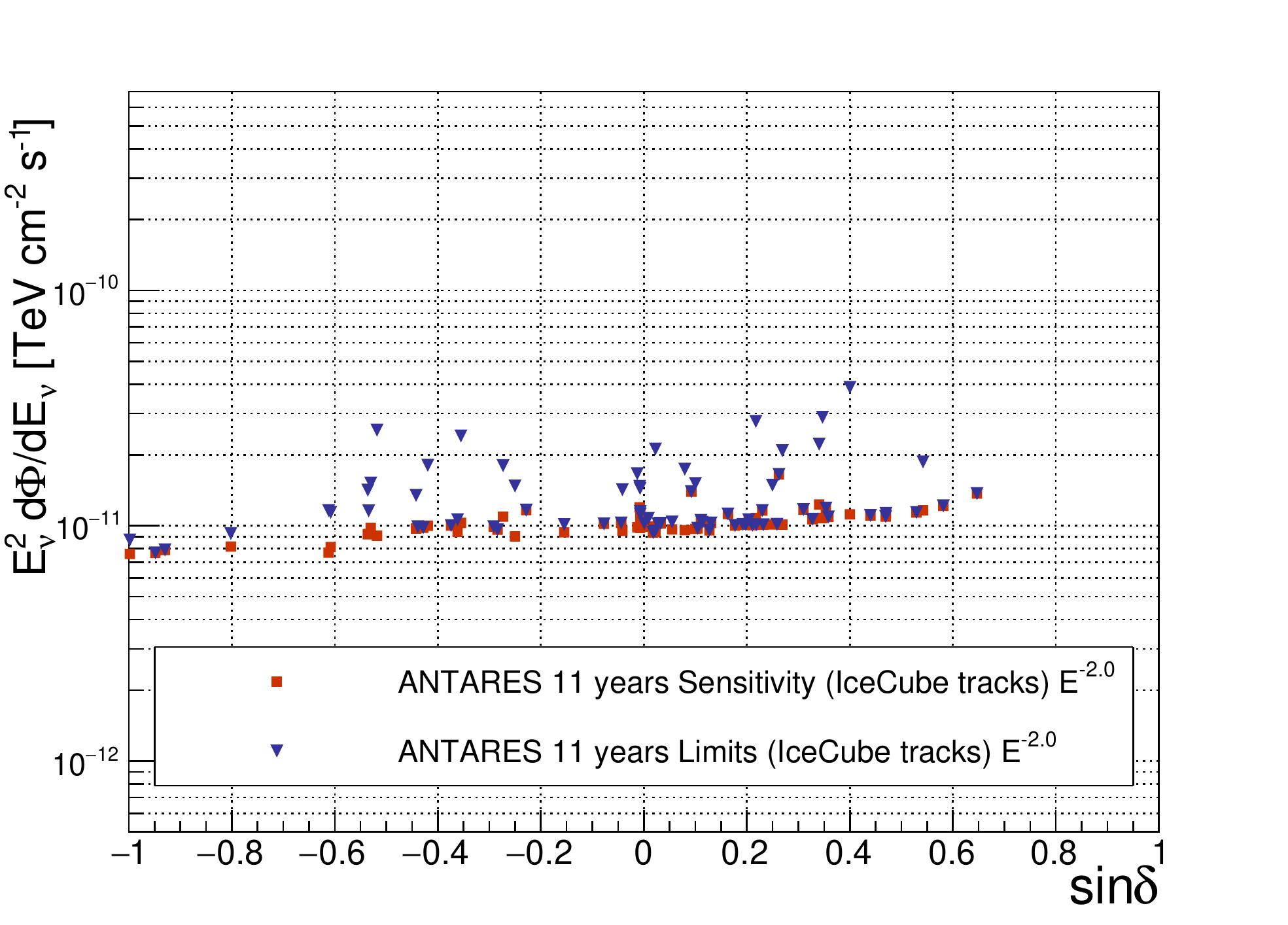}
     \put (11,60){\color{black} PRELIMINARY}
    \end{overpic}
\caption{Left: $90\,\%$ C.L. upper limits on the signal flux from the investigated astrophysical candidates (blue dots) as a function of the source declination for an $E^{-2.0}$ spectrum. The orange lines show the median sensitivity of this analysis for an $E^{-\gamma}$ spectrum, with $\gamma = 2.0$ (solid line) and $\gamma = 3.0$ (dashed line). Right: $90\,\%$ C.L. upper limits (blue triangles) and sensitivities (orange squares) for the investigated IceCube tracks as a function of the source declination for a $E^{-2.0}$ spectrum.}
\label{fig:Limits}
\end{figure*}

\textit {Search over astrophysical objects.} The candidate list used in \cite{lastPSant}, containing neutrino source candidates both from Galactic and extra-Galactic origin, is updated including five new sources reported in the TeVCat catalog \cite{tevcat} and detected after January 2016. The list of the 112 analysed candidates, together with the obtained results at each location, is reported in Table~\ref{tab:LimitsFix}. The most signal-like cluster is found at the location of HESSJ0632+057, at equatorial coordinates $(\alpha,\delta) = (98.24^\circ, 5.81^\circ)$, with a pre-trial p-value of 0.15\%, corresponding to a post-trial significance of $1.4\sigma$. The cluster contains 11(1) tracks within $5^\circ (1^\circ)$ and 2 shower events within $5^\circ$ around the source candidate, as shown in Figure~\ref{fig:Clusters}--middle.
    The 90\% C.L. limits for this search (assuming an $E^{-2.0}$ spectrum), calculated with the Neyman method \cite{neyman}, are shown in Figure~\ref{fig:Limits}-left as a function of the declination. For the special case of Eta Carinae, in addition to the $E^{-2.0}$ hypothesis, the upper limits have been also computed for three different neutrino spectra featuring an energy cut-off ($E_\mathrm{cut}=100$ TeV, 300 TeV and 1 PeV), following the predictions of \cite{Razzaque}. The resulting upper limits, expressed in multiples of the reference flux are: $\Phi^{90\%}= 6.9 \times \Phi(\mathrm{E_{cut}}=100\, \mathrm{TeV})$, $\Phi^{90\%}= 3.6 \times \Phi(\mathrm{E_{cut}}=300\, \mathrm{TeV})$ and $\Phi^{90\%}= 2.1 \times \Phi(\mathrm{E_{cut}}=1\, \mathrm{PeV})$.

\setlength{\tabcolsep}{.3em}

\begin{table*}[t]
    \caption{\footnotesize List of analysed astrophysical objects. Reported are the source's name, equatorial coordinates, best-fit number of signal events, pre-trial p-value and 90\% C.L. upper limits on the flux normalization factor for a $E^{-2.0}$ spectrum, $\Phi^{90\%}_{E^{-2.0}}$ (in units of $10^{-8} \ \rm{GeV cm^{-2} s^{-1}}$). Dashes (-) in the fitted number of source events and pre-trial p-value indicate sources with null observations.
  \medskip}
    \label{tab:LimitsFix}
    \resizebox{\linewidth}{!}{
        \begin{minipage}{1.2\textwidth}
        \centering
 {\def\arraystretch{0.8}
    \footnotesize
    \begin{tabular}{l crrcc | l crrcc}
   
             Name & $\delta [\si{\degree}]$   &    $\alpha [\si{\degree}]$    & $\hat{\mu}_\mathrm{sig}$ & p-value & $\Phi^{90\%}_{E^{-2.0}}$ &  Name & $\delta [\si{\degree}]$   &    $\alpha [\si{\degree}]$    & $\hat{\mu}_\mathrm{sig}$ & p-value & $\Phi^{90\%}_{E^{-2.0}}$  \\
\hline
LHA120-N-157B & -69.16 & 84.43 & -- & -- & 0.53 & HESSJ1837-069 & -6.95 & 279.41 & -- & -- & 0.93 \\ 
HESSJ1356-645 & -64.50 & 209.00 & 0.2 & 0.67 & 0.75 & 2HWCJ1309-054 & -5.49 & 197.31 & -- & -- & 0.83 \\ 
PSRB1259-63 & -63.83 & 195.70 & -- & -- & 0.53 & 3C279 & -5.79 & 194.05 & 0.8 & 0.10 & 1.35 \\ 
HESSJ1303-631 & -63.20 & 195.75 & -- & -- & 0.55 & 2HWCJ1852+013* & 1.38 & 283.01 & -- & -- & 0.84 \\ 
RCW86 & -62.48 & 220.68 & -- & -- & 0.53 & W44 & 1.38 & 284.04 & -- & -- & 0.84 \\ 
HESSJ1507-622 & -62.34 & 226.72 & -- & -- & 0.53 & PKS0736+017 & 1.62 & 114.83 & -- & -- & 0.94 \\ 
HESSJ1458-608 & -60.88 & 224.54 & 1.1 & 0.13 & 0.90 & RGBJ0152+017 & 1.79 & 28.17 & -- & -- & 0.84 \\ 
ESO139-G12 & -59.94 & 264.41 & -- & -- & 0.59 & 2HWCJ1902+048* & 4.86 & 285.51 & -- & -- & 0.85 \\ 
Eta Carinae & -59.68 & 161.27 & -- & -- & 1.00 & SS433 & 4.98 & 287.96 & -- & -- & 0.85 \\ 
MSH15-52 & -59.16 & 228.53 & -- & -- & 0.54 & HESSJ0632+057 & 5.81 & 98.24 & 2.7 & 0.0015 & 2.61 \\ 
HESSJ1503-582 & -58.74 & 226.46 & -- & -- & 0.54 & MGROJ1908+06 & 6.27 & 286.99 & -- & -- & 0.85 \\ 
HESSJ1023-575 & -57.76 & 155.83 & 1.3 & 0.12 & 0.93 & 2HWCJ1829+070 & 7.03 & 277.34 & -- & -- & 0.85 \\ 
CirX-1 & -57.17 & 230.17 & -- & -- & 0.57 & B1030+074 & 7.19 & 158.39 & -- & -- & 0.85 \\ 
SNRG327.1-01.1 & -55.08 & 238.65 & -- & -- & 0.58 & 2HWCJ1907+084* & 8.50 & 286.79 & -- & -- & 0.87 \\ 
HESSJ1614-518 & -51.82 & 243.58 & 0.8 & 0.18 & 0.82 & OT081 & 9.65 & 267.89 & -- & -- & 1.19 \\ 
HESSJ1616-508 & -50.97 & 243.97 & 0.6 & 0.18 & 0.81 & HESSJ1912+101 & 10.15 & 288.21 & -- & -- & 0.86 \\ 
PKS2005-489 & -48.82 & 302.37 & 0.2 & 0.76 & 0.74 & PKS1502+106 & 10.52 & 226.10 & -- & -- & 0.86 \\ 
GX339-4 & -48.79 & 255.70 & -- & -- & 0.55 & RBS0723 & 11.56 & 131.80 & -- & -- & 0.86 \\ 
HESSJ1632-478 & -47.82 & 248.04 & 1.0 & 0.15 & 0.86 & 2HWCJ1914+117 & 11.72 & 288.68 & -- & -- & 0.86 \\ 
RXJ0852.0-4622 & -46.37 & 133.00 & -- & -- & 0.54 & 2HWCJ1921+131 & 13.13 & 290.30 & -- & -- & 0.86 \\ 
HESSJ1641-463 & -46.30 & 250.26 & 1.3 & 0.099 & 0.94 & W51C & 14.19 & 290.75 & -- & -- & 0.86 \\ 
VelaX & -45.60 & 128.75 & -- & -- & 0.54 & 2HWCJ0700+143 & 14.32 & 105.12 & -- & -- & 1.24 \\ 
PKS0537-441 & -44.08 & 84.71 & 0.4 & 0.20 & 0.80 & VERJ0648+152 & 15.27 & 102.20 & -- & -- & 1.23 \\ 
CentaurusA & -43.02 & 201.36 & -- & -- & 0.56 & 2HWCJ0819+157 & 15.79 & 124.98 & -- & -- & 0.87 \\ 
PKS1424-418 & -42.10 & 216.98 & 1.0 & 0.13 & 0.88 & 3C454.3 & 16.15 & 343.50 & -- & -- & 0.88 \\ 
1ES2322-409 & -40.66 & 351.20 & -- & -- & 0.58 & PKS0235+164 & 16.61 & 39.66 & 1.9 & 0.062 & 1.75 \\ 
RXJ1713.7-3946 & -39.75 & 258.25 & -- & -- & 0.60 & Geminga & 17.77 & 98.47 & 0.8 & 0.14 & 1.49 \\ 
PKS1440-389 & -39.14 & 220.99 & 2.8 & 0.0060 & 1.61 & 2HWCJ1928+177 & 17.78 & 292.15 & -- & -- & 0.90 \\ 
PKS0426-380 & -37.93 & 67.17 & -- & -- & 0.61 & RGBJ2243+203 & 20.35 & 340.98 & -- & -- & 0.94 \\ 
PKS1454-354 & -35.67 & 224.36 & 1.2 & 0.097 & 1.13 & VERJ0521+211 & 21.21 & 80.44 & 1.0 & 0.13 & 1.53 \\ 
PKS0625-35 & -35.49 & 96.78 & -- & -- & 0.64 & 4C+21.35 & 21.38 & 186.23 & -- & -- & 0.95 \\ 
TXS1714-336 & -33.70 & 259.40 & 0.8 & 0.10 & 1.11 & Crab & 22.01 & 83.63 & -- & -- & 1.29 \\ 
SwiftJ1656.3-3302 & -33.04 & 254.07 & -- & -- & 0.86 & IC443 & 22.50 & 94.21 & -- & -- & 0.96 \\ 
PKS0548-322 & -32.27 & 87.67 & -- & -- & 0.69 & S20109+22 & 22.74 & 18.02 & -- & -- & 0.97 \\ 
H2356-309 & -30.63 & 359.78 & -- & -- & 0.71 & B1422+231 & 22.93 & 216.16 & -- & -- & 0.97 \\ 
PKS2155-304 & -30.22 & 329.72 & -- & -- & 0.70 & PKS1424+240 & 23.79 & 216.75 & -- & -- & 0.98 \\ 
HESSJ1741-302 & -30.20 & 265.25 & 0.6 & 0.14 & 1.10 & 2HWCJ1938+238 & 23.81 & 294.74 & -- & -- & 0.98 \\ 
PKS1622-297 & -29.90 & 246.50 & -- & -- & 0.70 & 2HWCJ1949+244 & 24.46 & 297.42 & -- & -- & 1.16 \\ 
GalacticCentre & -29.01 & 266.42 & 1.2 & 0.10 & 1.20 & MS1221.8+2452 & 24.61 & 186.10 & -- & -- & 0.99 \\ 
Terzan5 & -24.90 & 266.95 & -- & -- & 0.93 & PKS1441+25 & 25.03 & 220.99 & -- & -- & 1.00 \\ 
1ES1101-232 & -23.49 & 165.91 & -- & -- & 0.76 & 1ES0647+250 & 25.05 & 102.69 & 0.2 & 0.46 & 1.40 \\ 
PKS0454-234 & -23.43 & 74.27 & -- & -- & 0.75 & S31227+25 & 25.30 & 187.56 & -- & -- & 1.00 \\ 
W28 & -23.34 & 270.43 & 0.8 & 0.096 & 1.27 & WComae & 28.23 & 185.38 & -- & -- & 1.04 \\ 
PKS1830-211 & -21.07 & 278.42 & -- & -- & 0.76 & 2HWCJ1955+285 & 28.59 & 298.83 & -- & -- & 1.04 \\ 
SNRG015.4+00.1 & -15.47 & 274.52 & -- & -- & 0.92 & TON0599 & 29.24 & 179.88 & -- & -- & 1.05 \\ 
LS5039 & -14.83 & 276.56 & -- & -- & 1.04 & 2HWCJ1953+294 & 29.48 & 298.26 & -- & -- & 1.05 \\ 
QSO1730-130 & -13.10 & 263.30 & -- & -- & 0.80 & 1ES1215+303 & 30.10 & 184.45 & -- & -- & 1.06 \\ 
HESSJ1826-130 & -13.01 & 276.51 & -- & -- & 0.87 & 1ES1218+304 & 30.19 & 185.36 & -- & -- & 1.06 \\ 
HESSJ1813-126 & -12.68 & 273.34 & -- & -- & 0.80 & HESSJ1746-308 & 30.84 & 266.57 & -- & -- & 1.07 \\ 
1ES0347-121 & -11.99 & 57.35 & -- & -- & 0.83 & 2HWCJ1040+308 & 30.87 & 160.22 & -- & -- & 1.19 \\ 
PKS0727-11 & -11.70 & 112.58 & 1.2 & 0.076 & 1.43 & 2HWCJ2006+341 & 34.18 & 301.55 & -- & -- & 1.10 \\ 
HESSJ1828-099 & -9.99 & 277.24 & 1.6 & 0.077 & 1.45 & S30218+35 & 35.94 & 35.27 & 0.8 & 0.099 & 1.92 \\ 
HESSJ1831-098 & -9.90 & 277.85 & -- & -- & 0.81 & MGROJ2019+37 & 36.83 & 304.64 & 0.5 & 0.15 & 1.73 \\ 
HESSJ1834-087 & -8.76 & 278.69 & -- & -- & 0.81 & MilagroDiffuse & 38.00 & 305.00 & 0.4 & 0.15 & 1.73 \\ 
PKS1406-076 & -7.90 & 212.20 & -- & -- & 0.82 & Markarian421 & 38.19 & 166.08 & -- & -- & 1.22 \\ 
QSO2022-077 & -7.60 & 306.40 & 1.4 & 0.047 & 1.57 & B32247+381 & 38.43 & 342.53 & -- & -- & 1.22 \\ 
        \end{tabular}
      }
      \end{minipage}
      
    }
  
\end{table*}
    
\textit {Search for steady emission from the direction of the IceCube tracks.} A separate candidate list search is performed to analyse a total of 75 IceCube neutrino candidates classified as tracks. The list of investigated candidates includes the 20 events from the IceCube ``High-Energy Starting Events" (HESE) sample \cite{IC3years, IC4yproc, IC6yproc} and the 34 events from the IceCube ``Extremely High-Energy Events" (EHE) sample \cite{IC-VHE-29, IC-VHE-36} investigated in the ANTARES time-dependent search \cite{timecorr} described below. In addition, a total of 21 IceCube AMON alerts (12 HESE \cite{AMONhese} and 9 EHE \cite{AMONehe}) are included. The list in Table~\ref{tab:LimitsFixIC} includes the values of the estimated angular error of each candidate, provided by the IceCube Collaboration. Given the non-negligible angular uncertainty of the IceCube candidates, the position in the sky of the fitted source is left free to vary around the position of the IceCube event within a cone with opening angle twice as large as its estimated angular error. The results in Table~\ref{tab:LimitsFixIC} show that the IceCube track candidate with the largest excess is the EHE event with ID 3. The fitted cluster is located at $(\alpha,\delta) = (343.7^\circ, 23.6^\circ)$, which is at a distance of 0.2$^\circ$ from the original EHE track at $(\alpha,\delta) = (343.55^\circ, 23.78^\circ)$, and coincident with the most significant cluster found in the full-sky search (see Section \ref{fullsky}). The trial-corrected significance of the source is 1.5\% (2.4$\sigma$). Figure~\ref{fig:Limits}--right shows the 90\% C.L. sensitivities and limits on the neutrino flux from the investigated IceCube candidates as a function of the declination.
    
\textit {Search for transient emission from the direction of the IceCube tracks.} In this search, the 54 selected IceCube tracks from the HESE and EHE samples are treated as potential transient neutrino sources. In contrast to time-integrated searches, the information of the neutrino arrival times is exploited to enhance the discovery potential. When dealing with transient emissions, the background of atmospheric neutrinos can be significantly reduced by limiting the search to a small time window around the source flare. A generic Gaussian time profile for the signal emission is assumed, $\mathcal{S}^{\rm time}(t_i) = \frac{1}{\sqrt{2\pi}\sigma_t}e^{(-\frac{(t_i - t_{IC})^2}{2\sigma_t^2})}$,  with $t_i$ being the detection time of the ANTARES event $i$, $t_{IC}$ the observation time of the considered IceCube candidate, and $\sigma_t$ the unknown flare duration, free to vary in the likelihood maximisation between 0.1 and 120 days.
The most significant cluster of this search is found at the location of the EHE event with ID 15, with a best-fit flare duration $\hat{\sigma}_t = 120$ days and a pre-trial p-value of 3.7\%, corresponding to a post-trial significance of 90\%. A summary of the results, in terms of best-fit flare duration $\hat{\sigma}_t$ and upper limits on the fluence, is reported in Table~\ref{tab:LimitsFixIC}. For details on the calculation of the fluence, refer to  \cite{timecorr}.

\begin{table*}[t]
    \caption{\footnotesize List of analysed IceCube tracks. Reported are the candidate's sample and ID, estimated angular error, results of the time-integrated analysis (best-fit equatorial coordinates, best-fit number of signal events, pre-trial p-value and 90\% C.L. upper limits on the flux normalization factor for an $E^{-2.0}$ spectrum, $\Phi^{90\%}_{E^{-2.0}}$ in units of $10^{-8} \ \rm{GeV cm^{-2} s^{-1}}$) and results of the time-dependent analysis (best-fit flare duration $\hat{\sigma}_t$ and $90\,\%$ C.L. upper limits on the neutrino fluence in GeV cm$^{-2}$ for the energy spectra: $E^{-2.0}$ and $E^{-2.5}$). Dashes (-) in the fitted number of source events and pre-trial p-value indicate cases with null observations in the time-integrated analysis. Dashes in the fitted flare duration  indicate sources with a null number of fitted signal events in the time-dependent analysis.  \medskip}
    \label{tab:LimitsFixIC}
    \resizebox{\linewidth}{!}{
        \begin{minipage}{1.6\textwidth}
        \centering
 {\def\arraystretch{0.9}
    \footnotesize
    \begin{tabular}{lcc | llccc | ccc | lcc | llccc | ccc}
   
             Sample & ID & $\beta [\si{\degree}]$ & $\hat{\delta} [\si{\degree}]$   &    $\hat{\alpha} [\si{\degree}]$  & $\hat{\mu}_\mathrm{sig}$ & p-value & $\Phi^{90\%}_{E^{-2.0}}$ & $\hat{\sigma}_t [days]$ & $\mathcal{F}^{90\%}_{E^{-2.0}}$ & $\mathcal{F}^{90\%}_{E^{-2.5}}$ &             Sample & ID & $\beta [\si{\degree}]$ & $\hat{\delta} [\si{\degree}]$   &    $\hat{\alpha} [\si{\degree}]$  & $\hat{\mu}_\mathrm{sig}$ & p-value & $\Phi^{90\%}_{E^{-2.0}}$ & $\hat{\sigma}_t [days]$ & $\mathcal{F}^{90\%}_{E^{-2.0}}$ & $\mathcal{F}^{90\%}_{E^{-2.5}}$ \\
\hline
 HESE & 3 & 1.4 & -29.9 & 130.1 & 6.6 & 0.000012 & 2.55 & 2.9 & 12.69 & 26.94 &  & 20 & 1.0 & 28.0 & 167.0 & -- & -- & 1.13 & -- & 15.61 & 29.79 \\ 
 & 5 & 1.2 & 1.5 & 112.7 & 2.8 & 0.032 & 1.15 & 120 & 18.86 & 46.75 &  & 21 & 1.0 & 14.5 & 91.2 & 1.0 & 0.029 & 1.49 & -- & 13.42 & 30.02 \\ 
 & 8 & 1.3 & -22.0 & 184.0 & 2.2 & 0.020 & 1.06 & 120 & 20.68 & 55.84 &  & 22 & 1.0 & -4.4 & 224.6 & 1.3 & 0.079 & 1.02 & 120 & 20.02 & 47.21 \\ 
 & 13 & 1.2 & 41.7 & 67.5 & -- & -- & 1.37 & 120 & 20.75 & 41.94 &  & 23 & 1.0 & 9.2 & 32.5 & 0.5 & 0.16 & 1.00 & 120 & 22.93 & 53.95 \\ 
 & 18 & 1.3 & -23.4 & 346.5 & 1.8 & 0.0020 & 1.81 & -- & 12.1 & 28.04 &  & 24 & 1.0 & 32.3 & 295.5 & 1.8 & 0.016 & 1.87 & 19.6 & 20.85 & 41.02 \\ 
 & 23 & 1.9 & -14.4 & 209.8 & 1.8 & 0.019 & 1.17 & 120 & 13.91 & 33.07 &  & 25 & 1.1 & 15.8 & 350.0 & 1.5 & 0.097 & 1.17 & -- & 13.62 & 29.3 \\ 
 & 28 & 1.3 & -71.8 & 162.9 & 1.4 & 0.041 & 0.77 & 120 & 7.87 & 20.37 &  & 26 & 1.0 & 1.6 & 104.5 & 2.6 & 0.0030 & 2.12 & 120 & 24.26 & 62.82 \\ 
 & 37 & 1.2 & 20.1 & 169.5 & -- & -- & 1.19 & -- & 14.27 & 30.33 &  & 27 & 1.0 & 12.9 & 109.0 & -- & -- & 1.01 & -- & 12.9 & 28.96 \\ 
 & 43 & 1.2 & -21.7 & 208.6 & -- & -- & 1.00 & 26.0 & 10.5 & 24.24 &  & 28 & 1.0 & 5.8 & 99.0 & 2.2 & 0.0098 & 1.74 & -- & 12.46 & 27.09 \\ 
 & 44 & 1.2 & -1.4 & 336.3 & 1.0 & 0.037 & 1.03 & 120 & 18.99 & 47.36 &  & 29 & 1.0 & 12.2 & 91.0 & -- & -- & 1.01 & -- & 12.89 & 28.39 \\ 
 & 45 & 1.2 & -85.2 & 241.4 & 1.9 & 0.011 & 0.87 & 64.3 & 8.46 & 20.98 &  & 30 & 1.0 & 25.6 & 324.0 & 1.0 & 0.13 & 1.11 & 114.2 & 24.14 & 53.4 \\ 
 & 53 & 1.2 & -35.9 & 240.5 & 2.9 & 0.0081 & 1.16 & 120 & 11.61 & 27.56 &  & 31 & 1.0 & 5.5 & 327.7 & 1.1 & 0.13 & 0.98 & -- & 12.4 & 25.83 \\ 
 & 58 & 1.3 & -34.8 & 101.9 & 1.8 & 0.0052 & 1.42 & 18.4 & 14.29 & 30.78 &  & 32 & 1.0 & 29.0 & 136.0 & 0.8 & 0.098 & 1.09 & 118.9 & 18.97 & 36.77 \\ 
 & 61 & 1.2 & -18.6 & 56.5 & 1.3 & 0.062 & 0.96 & -- & 11.5 & 24 &  & 33 & 1.5 & 18.4 & 200.5 & 1.2 & 0.0030 & 2.22 & -- & 13.8 & 30.75 \\ 
 & 62 & 1.3 & 11.4 & 188.2 & 0.8 & 0.083 & 1.16 & -- & 13.14 & 28.67 &  & 34 & 1.0 & 11.1 & 76.2 & -- & -- & 1.01 & -- & 13.29 & 28.93 \\ 
 & 63 & 1.2 & 4.4 & 158.4 & 1.6 & 0.058 & 1.03 & 120 & 13.02 & 27.69 &  & 35 & 1.0 & 16.6 & 152.5 & 2.1 & 0.0047 & 2.09 & 120 & 24.58 & 60.38 \\ 
 & 71 & 1.2 & -18.9 & 81.2 & 4.4 & 0.00022 & 2.41 & 120 & 23.95 & 61.21 & AMON HESE & 766165\_132518 & 1.3 & -38.8 & 64.5 & 1.2 & 0.014 & 1.14 &   &   &   \\ 
 & 76 & 1.2 & 0.0 & 238.5 & 0.6 & 0.12 & 1.12 & -- & 11.76 & 27.8 &  & 66688965\_132229 & 1.3 & -15.9 & 266.6 & 1.5 & 0.060 & 0.99 &   &   &   \\ 
 & 78 & 1.2 & 5.9 & 1.5 & -- & -- & 1.03 & -- & 12.42 & 27.07 &  & 36142391\_132143 & 1.3 & -55.7 & 129.6 & 1.1 & 0.014 & 0.93 &   &   &   \\ 
 & 82 & 1.2 & 7.9 & 243.0 & 1.7 & 0.046 & 1.12 & -- & 12.73 & 27.52 &  & 9759013\_132077 & 1.3 & -33.5 & 305.1 & 1.3 & 0.012 & 1.16 &   &   &   \\ 
 EHE & 1 & 1.0 & 2.4 & 28.0 & -- & -- & 0.95 & -- & 12.2 & 27.57 &  & 68269692\_131999 & 1.3 & -23.4 & 2.0 & 1.0 & 0.039 & 0.98 &   &   &   \\ 
 & 2 & 1.0 & 12.9 & 296.3 & 0.9 & 0.084 & 1.06 & 120 & 25.88 & 64.99 &  & 66412090\_131680 & 1.3 & -69.8 & 180.1 & 1.9 & 0.066 & 0.79 &   &   &   \\ 
 & 3 & 1.1 & 23.6 & 343.7 & 4.8 & 0.0000015 & 3.87 & 120 & 27.35 & 61.56 &  & 56068624\_130126 & 1.3 & -17.5 & 162.5 & 2.2 & 0.0043 & 1.80 &   &   &   \\ 
 & 5 & 1.0 & 20.0 & 309.0 & -- & -- & 1.09 & -- & 14.86 & 30.6 &  & 32674593\_129474 & 1.3 & -27.9 & 223.0 & 1.7 & 0.029 & 0.99 &   &   &   \\ 
 & 6 & 4.4 & 14.0 & 248.0 & 2.0 & 0.0025 & 1.65 & 120 & 19.88 & 50.08 &  & 65274589\_129281 & 1.3 & -26.3 & 307.9 & 2.0 & 0.010 & 1.35 &   &   &   \\ 
 & 7 & 1.0 & 14.4 & 267.5 & 0.4 & 0.21 & 1.01 & 120 & 15.29 & 33.56 &  & 38561326\_128672 & 1.1 & 11.3 & 39.5 & 2.1 & 0.010 & 2.78 &   &   &   \\ 
 & 8 & 1.0 & 10.1 & 329.5 & 1.1 & 0.062 & 1.01 & 120 & 18.59 & 43.31 &  & 58537957\_128340 & 1.5 & -29.9 & 199.5 & 2.8 & 0.0037 & 1.52 &   &   &   \\ 
 & 9 & 1.0 & 1.0 & 90.4 & 1.0 & 0.045 & 1.07 & 120 & 12.65 & 29 &  & 6888376\_128290 & 1.3 & -0.4 & 213.5 & 3.2 & 0.015 & 1.47 &   &   &   \\ 
 & 10 & 1.0 & 4.0 & 285.1 & 0.9 & 0.055 & 1.04 & 26.0 & 18.77 & 43.95 & AMON EHE & 42419327\_132508 & 1.0 & 6.4 & 118.5 & -- & -- & 1.05 &   &   &   \\ 
 & 11 & 1.0 & 1.0 & 310.0 & -- & -- & 0.94 & -- & 12.32 & 27.68 &  & 53411354\_131653 & 1.0 & -8.6 & 271.5 & 1.1 & 0.048 & 1.01 &   &   &   \\ 
 & 12 & 1.0 & 21.6 & 234.7 & 2.3 & 0.00040 & 2.89 & -- & 14.76 & 32.08 &  & 34507973\_131475 & 1.0 & -1.0 & 148.0 & 0.7 & 0.15 & 1.03 &   &   &   \\ 
 & 13 & 1.0 & 35.1 & 273.0 & 0.7 & 0.11 & 1.22 & -- & 18.34 & 34.86 &  & 17569642\_130214 & 1.0 & 7.3 & 340.5 & 0.8 & 0.093 & 0.96 &   &   &   \\ 
 & 14 & 2.1 & 4.6 & 317.0 & 2.8 & 0.0098 & 1.40 & 6.8 & 15.21 & 35.2 &  & 50579430\_130033 & 1.0 & 5.8 & 77.5 & 1.1 & 0.018 & 1.51 &   &   &   \\ 
 & 15 & 1.0 & 0.9 & 224.5 & 1.1 & 0.10 & 1.02 & 120 & 28.65 & 75.21 &  & 80305071\_129307 & 1.0 & -14.5 & 98.0 & 1.1 & 0.014 & 1.48 &   &   &   \\ 
 & 16 & 1.0 & 18.5 & 37.5 & 1.4 & 0.10 & 1.07 & -- & 13.91 & 31.22 &  & 80127519\_128906 & 1.0 & 14.1 & 46.1 & 0.7 & 0.17 & 1.01 &   &   &   \\ 
 & 17 & 1.0 & 33.0 & 200.4 & -- & -- & 1.14 & 120 & 22.82 & 45.07 &  & 26552458\_128311 & 1.0 & -2.0 & 123.7 & 2.0 & 0.0094 & 1.66 &   &   &   \\ 
 & 18 & 1.0 & 1.2 & 328.2 & 1.3 & 0.088 & 1.02 & 120 & 12.07 & 27.7 &  & 6888376\_128290 & 1.0 & -0.2 & 213.4 & 3.2 & 0.016 & 1.44 &   &   &   \\ 
 & 19 & 1.0 & -1.4 & 204.5 & 1.3 & 0.018 & 1.42 & 98.8 & 18.82 & 48.1 &  &  &  &  &  &  \\

        \end{tabular}
      }
      \end{minipage}
      
    }
  
\end{table*}

\subsection{TXS 0506+056}\label{tx}

On 22 September 2017, a high-energy neutrino-induced muon track, with high probability of being of astrophysical origin, was detected by the IceCube Collaboration \cite{ICTXS}. The neutrino-candidate event, IC170922A, was selected by the EHE online event filter and reported through a Gamma-ray Coordinates Network (GCN) Circular \cite{GCN}. The analysis of the ANTARES online data stream, promptly triggered by the IceCube alert, revealed no upgoing muon neutrino candidate within $3^\circ$ around the IC170922A direction and within $\pm 1$h centered on the event time \cite{ANTARESTXS}. Successively, it was determined that IC170922A was coincident in direction and time with a gamma-ray flare from the blazar TXS 0506+056 \cite{ICmultimessenger}, at equatorial coordinates $(\alpha,\delta) = (77.36^\circ, 5.69^\circ)$. Triggered by these findings, a time-dependent analysis was performed by the IceCube Collaboration \cite{ICTXS}, revealing a significant excess centered on December 13, 2014 and identified by two time-window shapes (one Gaussian-shaped and one box-shaped time window). Motivated by these observations, two different analyses were performed using ANTARES data to investigate the location of the blazar. These searches are presented below and in detail in \cite{ANTARESTXS}.

\begin{figure*}[ht]
\centering
\begin{overpic}[width=0.29\textwidth]{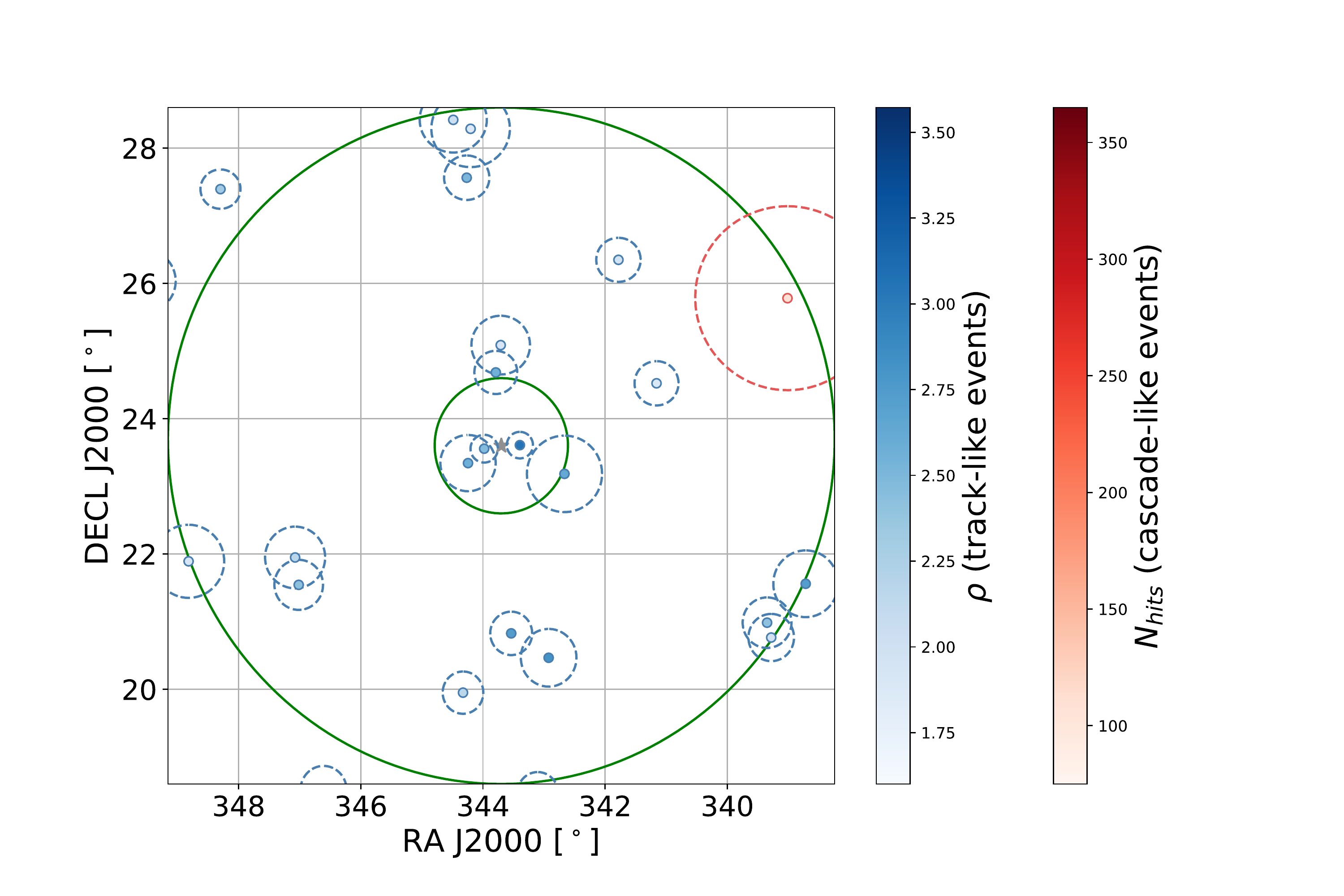}
     \put (29,92.7){\color{black} PRELIMINARY}
    \end{overpic}
    \begin{overpic}[width=0.29\textwidth]{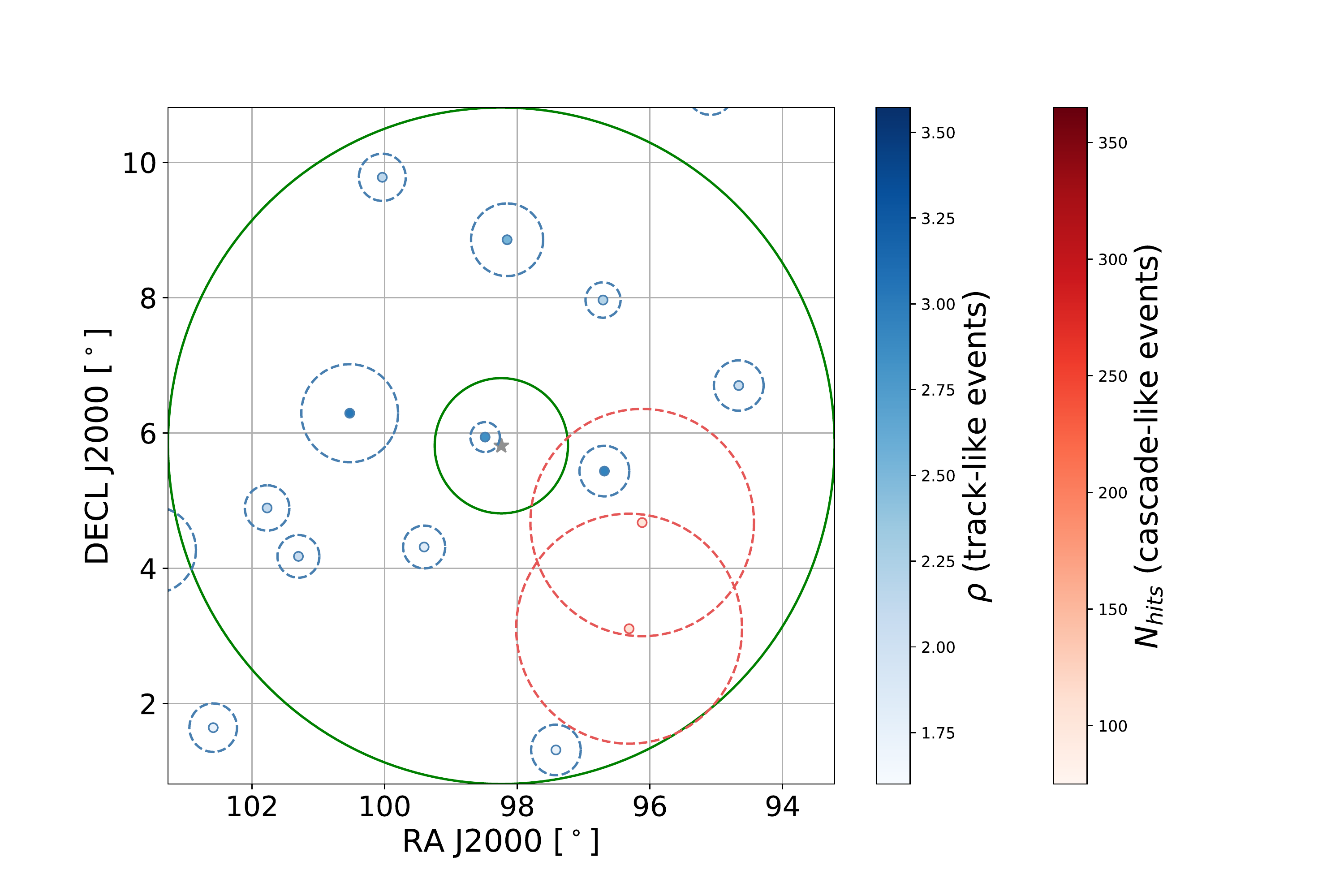}
     \put (28,92.7){\color{black} PRELIMINARY}
    \end{overpic}
        \begin{overpic}[width=0.29\textwidth]{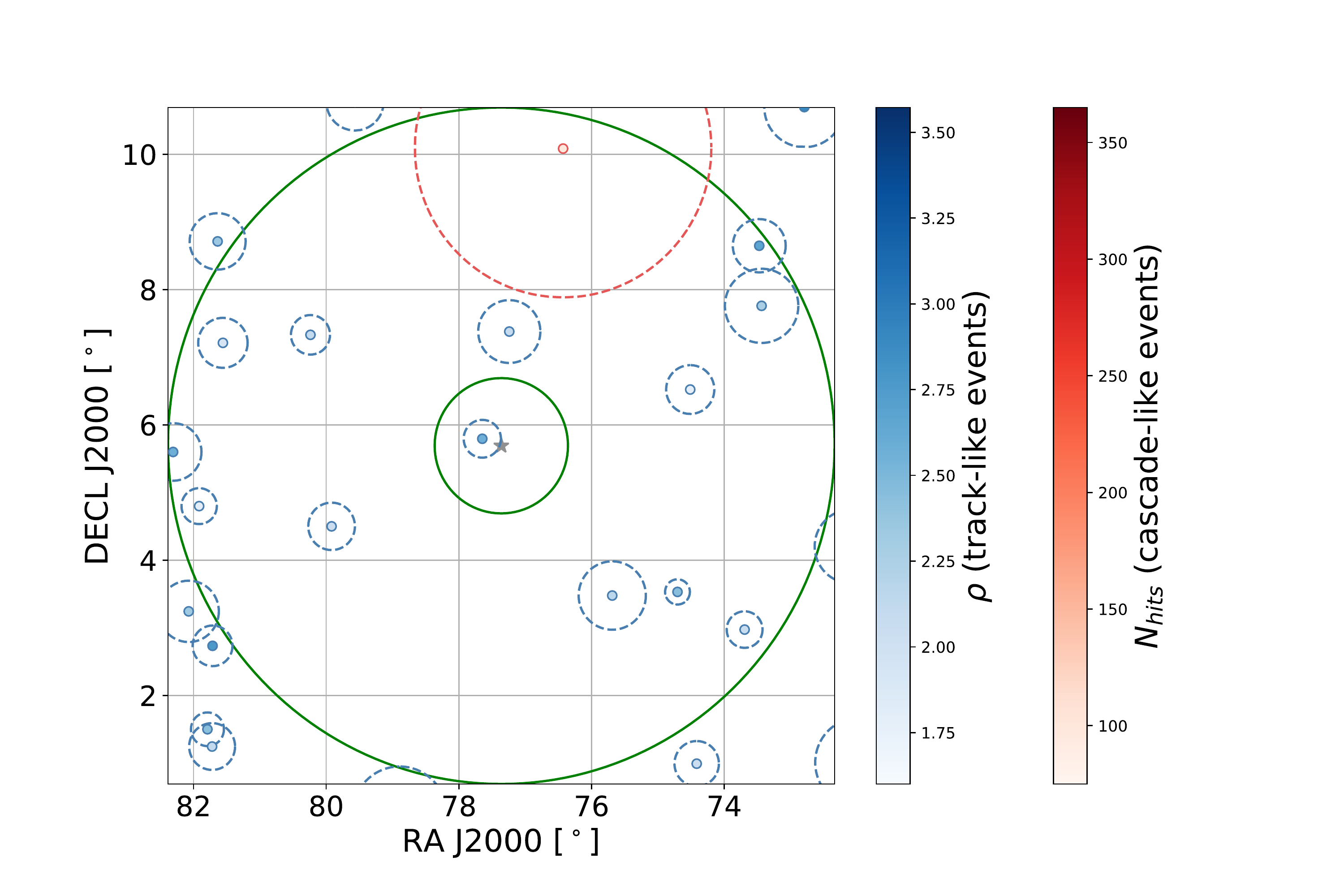}
     \put (22,92.7){\color{black}}
    \end{overpic}
        \begin{overpic}[width=0.11\textwidth]{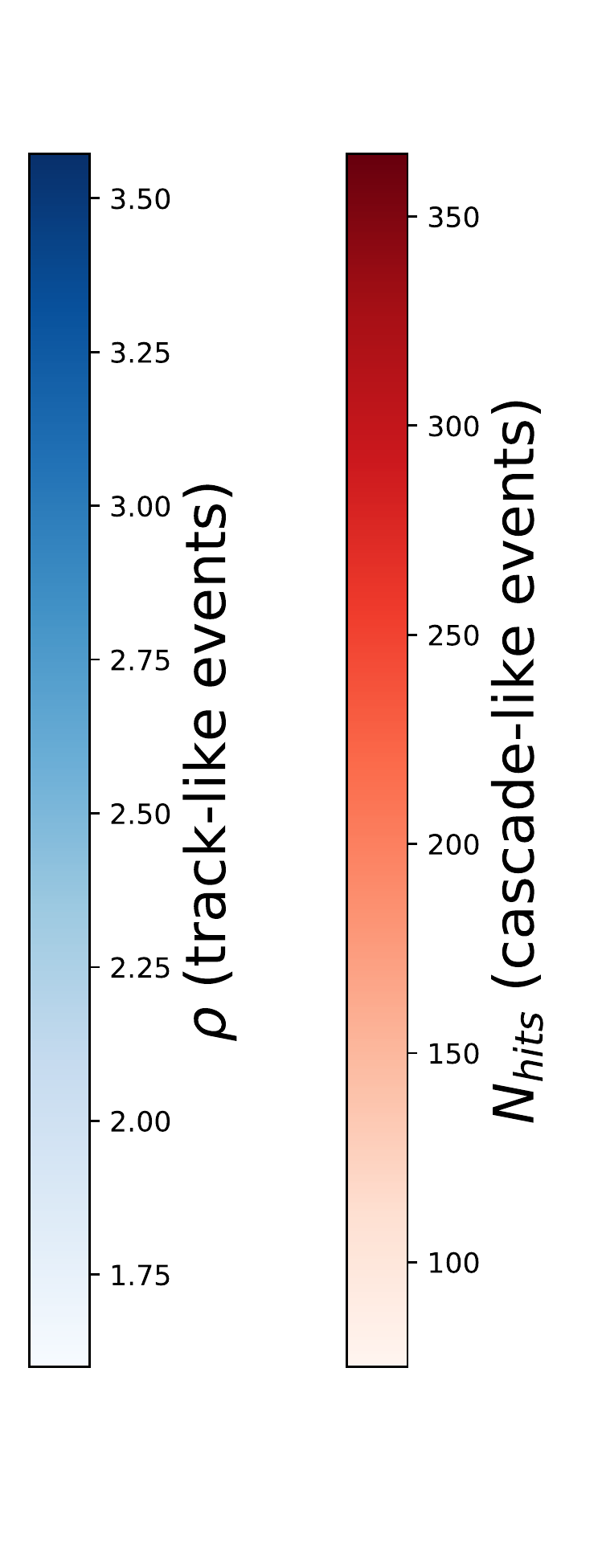}
     \put (11,10.7){\color{black}}
    \end{overpic}
\caption{Distribution of events in the ($\alpha$,  $\delta$) (RA, DEC) coordinates for the most significant cluster found in the full-sky search (left), the most significant cluster found in the candidate list search (HESSJ0632+057) (middle) and for the location of TXS 0506+056 (right). The inner (outer) green line depicts the one (five) degree distance from the position of the best-fit or known location, indicated as a gray star. The red points denote shower-like events, whereas the blue points indicate track-like events. The dashed circles around the events indicate the angular error estimate. Different tones of red and blue correspond to the values assumed by the energy estimators: the number of hits (shower-like events) and the $\rho$ parameter (track-like events) as shown in the legend. Refer to \cite{lastPSant} for further details on the energy estimators. }
\label{fig:Clusters}
\end{figure*}

\textit {Time-integrated search.} Searches for steady emission of neutrinos from the direction of TXS 0506+056 were carried out. The source was added to the list of 106 pre-selected sources analysed in \cite{lastPSant} and scrutinised using data recorded from the beginning of 2007 to the end of 2017. The cluster at the location of the blazar resulted to be the third most significant one out of the 107 investigated sources, with a number of fitted signal events $\mu_{sig} = 1.03$, a pre-trial p-value of 3.4\%  and a post-trial p-value of 87\% (for an unbroken power-law spectrum $E^{-2.0}$). One track-like event mostly influences the fit (see Figure~\ref{fig:Clusters}-right). It occurred on December 12, 2013 and is located within 1$\sigma$ from the source position. The value of the energy estimator, $\rho$, for this event is such that only 9\% of the neutrino candidates inducing a track have a larger value. 
From these null results, 90\% C.L. upper limits on the flux normalization factor at the energy of 100 TeV, $\Phi^{90\%}_{100\ \rm {TeV}}$ , assuming a steady neutrino source emitting with unbroken power-law spectra $E^{-2.0} (E^{-2.3}) [E^{-2.5}]$, were set to $1.6(1.4)[1.0] \times 10^{-18}$ GeV$^{-1}$ cm$^{-2}$ s$^{-1}$.

\textit {Time-dependent search.} A time-dependent analysis was performed assuming the two time profiles provided by the IceCube Collaboration. The first time profile is described by a Gaussian signal centred on MJD 57004 and with standard deviation $\sigma$ = 55.0 days. Neutrinos were searched for in a period $\pm 5\sigma$ wide, corresponding to 550 flaring days. The second one assumes a box-shaped flare starting at MJD 56937.81 and ending at MJD 57096.21, corresponding to 158.40 flaring days.
The search yielded no significant observation. Within $2^{\circ}$ from the source, 13 events were found in data, with 10 being the expected number of background events during the analysed period. None of the signal events was detected within either of the two considered flaring periods.
As no significant evidence of cosmic neutrino was observed, 90\% C.L. upper limits were derived for the neutrino flux. For the Gaussian-shaped period and unbroken energy spectra $E^{-2.0} (E^{-2.1}) [E^{-2.2}]$, the limits correspond to normalization factors of $\Phi^{90\%}_{100\ \mathrm{TeV}} = 4.6(4.4)[4.2] \times 10^{-18} \mathrm{GeV}^{-1}\mathrm{cm}^{-2}\mathrm{s}^{-1} $. The energy range containing the 5-95\% of the detectable flux is 2.0 (1.3) [1.0] TeV – 3.2 (1.6) [1.0] PeV. The limits on the flux normalization factors for box-shaped period are a factor 3.3 higher.

\section{Conclusions}\label{sec:cocl}
The results of various searches for point-like sources using events detected by the ANTARES telescope during 11 years of data taking have been presented. Searches for both steady and transient cosmic neutrino sources have been performed: a scan over the whole ANTARES visible sky, an investigation of 112 astrophysical candidates and 75 IceCube tracks, and a dedicated analysis of the direction of the TXS 0506+056 blazar. No significant evidence of cosmic neutrino sources has been found. The competitiveness of the results achieved demonstrates the huge potential of the new, cubic-kilometre scale, KM3NeT \cite{km3net}, which is expected to detect the neutrino flux reported by IceCube within a few months of operation and to make definite statements about a neutrino flux from several Galactic candidates. 

\vspace{15px}
\small{We gratefully acknowledge the financial support of the Ministerio de Ciencia, Innovación y Universidades: Programa Estatal de Generación de Conocmiento, ref. PGC2018-096663-B-C41 (MCIU/FEDER) and Severo Ochoa Centre of Excellence (MCIU), Spain.}

\bibliographystyle{ICRC}
%\bibliography{References}

\providecommand{\href}[2]{#2}\begingroup\raggedright\endgroup

\end{document}